\documentclass[aps]{revtex4}

\usepackage{amsmath}
\usepackage{amssymb}

\newcommand{\beq}{\begin{equation}}
\newcommand{\eeq}{\end{equation}}
\newcommand{\eps}{\epsilon}
\newcommand{\curl}{\nabla\times}
\newcommand{\vv}{\mathbf{v}}

\newcommand{\lap}{\Delta}
\newcommand{\ilap}{\Delta^{-1}}

\newcommand{\ben}{\begin{eqnarray}}
\newcommand{\een}{\end{eqnarray}}
\newcommand{\benn}{\begin{eqnarray*}}
\newcommand{\eenn}{\end{eqnarray*}}
\newcommand{\tn}{\tilde{n}}

\newcommand{\calL}{{\mathcal L}}
\newcommand{\hN}{\hat{N}}

\begin{document}

\title{Derivation of reduced two-dimensional fluid models\\ via Dirac's theory of constrained Hamiltonian systems}

\author{C. Chandre$^1$, E. Tassi$^1$, P.J. Morrison$^2$}
\affiliation{$^1$ Centre de Physique Th\'eorique, CNRS -- Aix-Marseille Universit\'e, Campus de Luminy, case 907, F-13288 Marseille cedex 09, France \\ $^2$ Institute for Fusion Studies and Department of Physics, The University of Texas at Austin, Austin, TX 78712-1060, USA}
\date{\today}
\baselineskip 24 pt

\begin{abstract}
We present a Hamiltonian derivation of a class of reduced plasma two-dimensional fluid models, an example being the Charney-Hasegawa-Mima equation. These models are obtained from the same parent Hamiltonian model, which consists of the ion momentum equation coupled to the continuity equation, by imposing dynamical constraints. It is shown that the Poisson bracket associated with these reduced models is the Dirac bracket obtained from the Poisson bracket of the parent model.

\bigskip

Key Words:  reduced fluid models, Dirac constraints, Hamiltonian, Poisson bracket
\end{abstract}

\maketitle



\section{Introduction}

Modelling plasma dynamics, with  kinetic or fluid approaches, often amounts to investigating  reduced, relatively tractable,  models  that capture the essential ingredients of the phenomenon under consideration while neglecting, for example, irrelevant spatial or temporal  scales.
Obtaining valuable reduced models provides a practical computational advantage when numerical simulations of a phenomenon have to be carried out. Reduced models are sometimes 
derived from parent models through a well-defined reduction procedure that 
amounts to  approximation directly at the level of the model equations, after 
having introduced some ordering based on physical arguments. Ideally, the reduced model should inherit some essential properties from the parent model. It has been recognized that the ideal part of such parent models possess a Hamiltonian (although noncanonical) character, consisting of a Hamiltonian functional, which can be identified as the total energy of the system, and a noncanonical Poisson bracket \cite{morr98}. Paradigmatic  examples are the  Vlasov-Maxwell equations~\cite{morr80b,mars82,bial84} and the ideal MHD equations~\cite{morr80a}, which both possess such a Hamiltonian structure. If one starts a derivation from a Hamiltonian parent model, then the final reduced model should also possess a Hamiltonian structure. If this were not the case,  some faulty dissipation would enter the reduced model,  and lead to qualitatively different interpretation of  physical behavior. For instance, numerical simulations of a reduced model with spurious dissipative terms might converge to attracting states, which would be ruled out if the Hamiltonian structure were present.  

When the physical arguments invoked to reduce the parent model take the form of constraints
on the dynamics, the method of Dirac brackets (see, e.g.\ Ref.~\cite{Dir50,Mar02,Sud74,Bhans76}), provides a systematic method for obtaining  a Hamiltonian reduced model from a Hamiltonian parent model. This  method is general  and  works  in the case of noncanonical Hamiltonian systems, which is the type of systems that arises when the models are formulated in terms of Eulerian variables, frequently used in plasma physics. Dirac brackets have proven to be useful for both finite and
infinite-dimensional \cite{nguy99,nguy01,nguy09}, for the derivation of balance models
in geophysical fluid dynamics  \cite{Sal88,vann02}, as a numerical (simulated annealing) method for calculating  vortex states \cite{morr05},  and also for describing the dynamics of fluids with free boundaries \cite{morr09}.  The purpose of the present  paper is to show how a class of reduced models, which will be Hamiltonian by construction, can be obtained from a Hamiltonian fluid parent model. Relevant examples of reduced models belonging to this class include the Charney-Hasegawa-Mima equation \cite{char71,hase77}, describing the propagation of drift waves in plasma,  and the Euler equation for an incompressible fluid. 

Here a slab geometry with Cartesian coordinates is adopted and  the dynamics of a plasma is confined to  a plane transverse to an imposed magnetic field.  Given a uniform and constant magnetic field ${\bf B}=B\hat{\bf z}$, the parent model  describes the ion fluid dynamics in terms of its density and velocity fields. This model has a noncanonical Hamiltonian form~\cite{morr80a}. The question addressed is how to derive 
reduced models describing the evolution of a few fields (for instance, the density or the  electrostatic potential), given some physical constraints dictated by experimental relevance, while preserving the Hamiltonian character of the parent model. In order to perform the reduction,  
two constraints are considered: First an incompressibility assumption on the ion fluid, and second a 
relation between the density and the electrostatic potential (or equivalently of the streamfunction of the incompressible part of the ion velocity field). This last constraint can eventually be ascribed 
to a relation that couples the ion and the electron fluid  through a quasi-neutrality assumption. We show using 
 Dirac brackets that the Poisson bracket of the reduced fluid models can be constructed and the noncanonical Hamiltonian structure of these reduced models are recovered in a systematic way.

The paper is organized as follows: In Sec.~\ref{sec:ion} we recall the Hamiltonian structure of the parent model and introduce new dynamical variables that are particularly convenient for the problem under consideration. In Sec.~\ref{sec:dirac} the essential elements of Dirac's theory of constrained Hamiltonian systems are briefly reviewed and the explicit derivation of the Dirac bracket for the reduced models is carried out. In Sec.~\ref{sec:hamilt} we focus on the specific examples of the Euler equation for an incompressible fluid and the Charney-Hasegawa-Mima equation. We also discuss here the model derived by Terry and Horton \cite{terr82}. Conclusions are drawn in Sec.~\ref{sec:concl}. 



\section{Expansion of the Ion fluid dynamics around equilibrium}
\label{sec:ion}

We start the derivation from a parent model with  two dynamical equations: one describing the transverse dynamics of the ion velocity field $\vv(x,y,t)$ and the other describing the dynamics of the ion density field $n(x,y,t)$:
\ben
&& \dot{\vv} +(\vv \cdot \nabla) \vv=-\nabla\varphi+\vv \times {\bf B},\label{eqn:s1}\\
&& \dot{n}=-\nabla\cdot (n\vv),\label{eqn:s2}
\een
where the dot indicates the partial derivative with respect to time $t$. We have used units such that the ion mass is $M=1$, its charge $e=1$,  and the amplitude of the magnetic field $B=1$.
The total energy of the ions, given by the sum of their kinetic energy plus the potential energy provided by the electrostatic potential $\varphi$, is a conserved quantity that is also a good candidate for the Hamiltonian of the system of Eqs.~(\ref{eqn:s1}-\ref{eqn:s2}), viz.
\beq
H[n,\mathbf{v}]=\int d^2 x \left[n\frac{v^2}{2}+n\varphi\right].
\label{psiext}
\eeq
The dynamics is determined by the Poisson bracket~\cite{morr80a,tass09}
\beq \label{br1}
\{F,G\}=-\int d^2 x \left[F_{\mathbf{v}}\cdot \nabla G_n - \nabla F_n\cdot G_{\mathbf{v}}-\left(\frac{\curl \bf{v}+\hat{\bf z}}{n}\right)\cdot F_{\mathbf{v}}\times G_{\mathbf{v}}\right],
\eeq
where we denote the functional derivatives of a given observable $F[n,\vv]$ by subscripts, i.e.\ $F_{\vv}=\delta F/\delta \vv$ and $F_n=\delta F/\delta n$.
In our context we assume that the electrostatic potential $\varphi$ is determined by the dynamics of the electrons which leads to a function $\varphi(n_e)$, where $n_e$ is the electron density. From the quasi-neutrality condition, $n=n_e$,  the Hamiltonian becomes 
\beq
H[n,\mathbf{v}]=\int d^2 x \left[n\frac{v^2}{2}+\psi(n)\right],
 \label{ham}
 \eeq
where $\psi'(n)=\varphi(n)$. We notice that for an external potential $\varphi$, we obtain $\psi(n)=n\varphi$ as in Eq.~(\ref{psiext}). Another example is obtained by neglecting the inertia of the electrons so that their density obeys a Boltzmann law $n_e=n_0 \exp\varphi$, where $n_0=n_0(x,y)=1-\lambda(x,y)$ is the electron density at equilibrium,  given by a constant plus a space dependent part that we will later assume to be small. Hence,   $\varphi(n)=\log(n/(1-\lambda))$ and consequently $\psi(n)=n(\log(n/(1-\lambda))-1)$. 

We perform the following change of variables~$(n,\vv)\mapsto (\tn,\phi,D)$ defined by
\benn
&& \tn=n,\\
&& \lap \phi=\hat{\bf z}\cdot\nabla\times \vv,\\
&& \lap D=\nabla\cdot \vv\,,
\eenn
where $\lap$ denotes the Laplacian. For simplicity, we use $n$ instead of $\tn$ in what follows.
In terms of the new variables $(n,\phi,D)$, Hamiltonian~(\ref{ham}) becomes
\begin{equation}
\label{eq:hamnv}
H[n,\phi,D]=\int d^2 x \left[n\left(\frac{|\nabla \phi|^2+|\nabla D|^2}{2}+[ \phi,D]\right)+\psi(n)\right],
\end{equation}
where $[f,g]=\hat{\bf z}\cdot \nabla f \times \nabla g$, and the bracket~(\ref{br1}) becomes
\benn
\{F,G\}&=&\int d^2 x \Big( F_n G_D - F_D G_n +\frac{\lap\phi +1}{n}\big( [\ilap F_\phi,\ilap G_\phi]
   \\
&& +  [\ilap F_D,\ilap G_D] +\nabla\ilap F_D\cdot \nabla \ilap G_\phi-\nabla\ilap F_\phi\cdot \nabla \ilap G_D\big)\Big)\,.
\eenn

We first assume that the variables evolve slowly with time, which is equivalent to adding a prefactor of  $1/\eps$  to  the Hamiltonian, and we introduce an $\eps$-ordering for the dynamical variables. The hypothesis is that the system of interest is near an equilibrium state whose spatial variations are of order $\eps$:
\benn
&& n=1+\eps n_1,\\
&& \phi=\eps \phi_1,\\
&& D=\eps D_1.
\eenn
The expansion of the Poisson bracket up to order $O(\eps^0)$ is given by 
\ben
\{F,G\}&=& \frac{1}{\eps^2}\int d^2x \left(F_{n_1} G_{D_1}-F_{D_1} G_{n_1}  -F_{D_1}\ilap G_{\phi_1} +
F_{\phi_1}\ilap G_{D_1} \right) \nonumber\\
&& -\frac{1}{\eps} \int d^2x \left( F_{\phi_1}\ilap \calL \ilap G_{\phi_1}+F_{D_1}\ilap \calL \ilap G_{D_1}-F_{D_1}\ilap \Lambda \ilap G_{\phi_1}+F_{\phi_1}\ilap \Lambda \ilap G_{D_1}\right),\label{eq:PB1}
\een
with the linear operators $\calL$ and $\Lambda$  defined by
\benn
&& \calL f=\left[\lap\phi_1-n_1,f\right],\\
&& \Lambda f= -\nabla\cdot \left((\lap\phi_1-n_1)\nabla f\right)\,,
\eenn
where we omit their dependence on $\lap\phi_1-n_1$.  Observe,  $\calL$ is anti-self-adjoint ($\calL^\dagger = -\calL$), while $\Lambda$ is self-adjoint ($\Lambda^\dagger=\Lambda$). 

In the next section, we impose constraints on the Poisson bracket~(\ref{eq:PB1}) and compute the associated Dirac bracket.



\section{Dirac brackets} \label{sec:dirac}

First we recall few basic facts about  Dirac brackets in infinite dimensions. If we impose $N$ Eulerian constraints $\Phi_\alpha({\bf x})=0$ for $\alpha=1,\ldots, N$ on a Hamiltonian system with a Hamiltonian $H$ and a Poisson bracket $\{\cdot,\cdot\}$, the Dirac bracket is obtained from the matrix $C$ defined by the Poisson bracket between the constraints
$$
C_{ab}({\bf x},{\bf x}')=\{\Phi_a({\bf x}),\Phi_b({\bf x}')\}\,, 
$$
where note $C_{\alpha \beta}({\bf x},{\bf x}')=-C_{\beta\alpha }({\bf x}',{\bf x})$.
The Dirac bracket is defined  by  
\begin{equation}
\label{eqn:dirac}
\{F,G\}_*=\{F,G\}-\int d^2x\int d^2x'\,  \{F,\Phi_\alpha({\bf x})\}C^{-1}_{\alpha \beta}({\bf x},{\bf x}')\{\Phi_\beta({\bf x}'),G\},
\end{equation}
where  the  $C^{-1}_{\alpha \beta}({\bf x},{\bf x}')$ are defined by
$$
\int d^2x' \, C^{-1}_{\alpha \beta}({\bf x},{\bf x}')C_{\beta\gamma}({\bf x}',{\bf x}'')=\int d^2x' \, 
C_{\alpha \beta}({\bf x},{\bf x}')C^{-1}_{\beta\gamma}({\bf x}',{\bf x}'')
=\delta_{\alpha \gamma}\delta({\bf x}-{\bf x}''),
$$
which implies $C^{-1}_{\alpha \beta}({\bf x},{\bf x}')=-C^{-1}_{\beta\alpha }({\bf x}',{\bf x})$.

Dirac obtained  (\ref{eqn:dirac}) from a modified Hamiltonian with Lagrange multipliers associated with each constraint:
$$
H'=H+\int d^2x \, \mu_{\alpha}({\bf x})\Phi_\alpha({\bf x}).
$$
The computation of the dynamical equation associated with this new Hamiltonian gives
$$
\{F,H'\}=\{F,H\}+\int d^2x \, \mu_\alpha \{F,\Phi_\alpha\}+\int d^2x\,  \Phi_\alpha\{F,\mu_\alpha\},
$$
which is equal to 
$$
\{F,H'\}\approx \{F,H\}_*\equiv \{F,H\}+\int d^2x \, \mu_\alpha \{F,\Phi_\alpha\} ,
$$
where the symbol $\approx$ means equality after  the constraints  are imposed.  The coefficients $\mu_\alpha$ are obtained by demanding that $\Phi_\alpha$ are Casimir invariants of the bracket $\{\cdot,\cdot\}_*$, which  leads to the definition of the bracket~(\ref{eqn:dirac}).  

We impose two constraints on the dynamics. The first one is incompressibility,  which translates here into
$$
\Phi_1({\bf x})=D_1=0,
$$ 
and the second one is an assumption relating  the electron density to the electrostatic field. This assumption takes the form
$$
\Phi_2({\bf x})=n_1-N(\phi_1)=0,
$$
where $N$ is, in general a nonlinear pseudo-differential function of $\phi_1$, i.e.\ a function of $\phi_1$ and its derivatives to arbitrary order. The functional derivative of $\Phi_2$ with respect to $\phi_1$ is given by
$$
\frac{\delta \Phi_2({\bf x})}{\delta \phi_1({\bf x}')}=-\hN^\dagger \delta({\bf x}'-{\bf x}),
$$
where $\hN$ is the Fr\'echet derivative of $N$ defined by 
$$
\hN\delta\phi_1=\left. \frac{d}{d\varepsilon} N(\phi_1+\varepsilon\delta \phi_1)\right|_{\varepsilon=0}.
$$
The Poisson brackets between the constraints are given by:
\begin{eqnarray}
	&& C_{11}({\bf x},{\bf x}')=-\frac{1}{\eps} \ilap \calL \ilap \delta({\bf x}-{\bf x}'),\label{eq:Cm11}\\
	&& C_{12}({\bf x},{\bf x}')=-\frac{1}{\eps^2}\left(1-\ilap \hN^\dagger+\eps \ilap\Lambda\ilap\hN^\dagger \right)\delta({\bf x}-{\bf x}'),\label{eq:Cm12}\\
	&& C_{21}({\bf x}',{\bf x})=\frac{1}{\eps^2}\left(1-\hN\ilap+\eps \hN\ilap\Lambda\ilap \right)\delta({\bf x}-{\bf x}'),\label{eq:Cm21}\\
	&& C_{22}({\bf x},{\bf x}')=-\frac{1}{\eps} \hN\ilap\calL\ilap\hN^\dagger \delta({\bf x}-{\bf x}').\label{eq:Cm22}
\end{eqnarray}
We take the convention that the linear operators act on the first mentioned independent variable, e.g.,   ${\hat{N}}\delta({\bf x}-{\bf x}')$ involves the derivatives with respect to ${\bf x}$, while ${\mathcal L}\delta({\bf x}'-{\bf x})$ involves the derivatives with respect to ${\bf x}'$. 
The elements of $C^{-1}$ are determined by the following four relations: The first two  determine $C_{11}^{-1}({\bf x},{\bf x}')$ and $C_{21}^{-1}({\bf x},{\bf x}')$:
\begin{eqnarray*}
	&& \int d^2x' \left[ C_{11}({\bf x},{\bf x}')C_{11}^{-1}({\bf x}',{\bf x}'')+C_{12}({\bf x},{\bf x}')C_{21}^{-1}({\bf x}',{\bf x}'')\right]=\delta({\bf x}-{\bf x}''),\\
	&& \int d^2x' \left[ C_{21}({\bf x},{\bf x}')C_{11}^{-1}({\bf x}',{\bf x}'')+C_{22}({\bf x},{\bf x}')C_{21}^{-1}({\bf x}',{\bf x}'')\right]=0\,,
\end{eqnarray*}
which upon using Eqs.~(\ref{eq:Cm11}-\ref{eq:Cm22})  become
\begin{eqnarray*}
	&& -\eps \ilap\calL\ilap C_{11}^{-1}-(1-\ilap\hN^\dagger+\eps\ilap\Lambda\ilap\hN^\dagger)C_{21}^{-1}=\eps^2 \delta({\bf x}-{\bf x}'),\\
	&& (1-\hN\ilap+\eps\hN\ilap\Lambda\ilap)C_{11}^{-1}-\eps \hN\ilap\calL\ilap\hN^\dagger C_{21}^{-1}=0.
\end{eqnarray*}
The expansion of the solution of the above equations, up to order $O(\eps^4)$, is given by
\begin{eqnarray*}
&& C_{11}^{-1}({\bf x},{\bf x}')=-\eps^3 (1-\hN\ilap)^{-1}\hN\ilap\calL\ilap\hN^\dagger(1-\ilap\hN^\dagger)^{-1}\delta({\bf x}-{\bf x}')+O(\eps^4),\\
&& C_{21}^{-1}({\bf x},{\bf x}')=-\eps^2 (1-\ilap\hN^\dagger)^{-1}\delta({\bf x}-{\bf x}')+\eps^3 (1-\ilap\hN^\dagger)^{-1}\ilap\Lambda\ilap\hN^\dagger (1-\ilap\hN^\dagger)^{-1} \delta({\bf x}-{\bf x}')+O(\eps^4).
\end{eqnarray*}
In the same way we obtain the expansions for $C_{22}^{-1}({\bf x},{\bf x}')$ and $C_{12}^{-1}({\bf x},{\bf x}')$:
\begin{eqnarray*}
	&& \int d^2x' \left[ C_{22}({\bf x},{\bf x}')C_{22}^{-1}({\bf x}',{\bf x}'')+C_{21}({\bf x},{\bf x}')C_{12}^{-1}({\bf x}',{\bf x}'')\right]=\delta({\bf x}-{\bf x}''),\\
	&& \int d^2x' \left[ C_{12}({\bf x},{\bf x}')C_{22}^{-1}({\bf x}',{\bf x}'')+C_{11}({\bf x},{\bf x}')C_{12}^{-1}({\bf x}',{\bf x}'')\right]=0,
\end{eqnarray*}
which upon using Eqs.~(\ref{eq:Cm11})--(\ref{eq:Cm22}) leads to
\begin{eqnarray*}
&& C_{22}^{-1}({\bf x},{\bf x}')=-\eps^3 (1-\ilap\hN^\dagger)^{-1}\ilap\calL\ilap(1-\hN\ilap)^{-1}\delta({\bf x}-{\bf x}')+O(\eps^4),\\
&& C_{12}^{-1}({\bf x},{\bf x}')=\eps^2 (1-\hN\ilap)^{-1}\delta({\bf x}-{\bf x}')-\eps^3(1-\hN\ilap)^{-1}\hN\ilap\Lambda\ilap(1-\hN\ilap)^{-1} \delta({\bf x}-{\bf x}')+O(\eps^4).
\end{eqnarray*}
Now, using the above and the following expressions:
\begin{eqnarray*}
	&& \{F,\Phi_1({\bf x})\}=\frac{1}{\eps^2}\left(F_{n_1}+\ilap F_{\phi_1}-\eps\ilap\Lambda\ilap F_{\phi_1}+\eps\ilap\calL\ilap F_{D_1}\right),\\ 
	&& \{F,\Phi_2({\bf x})\}=-\frac{1}{\eps^2}\left((1-\hN\ilap)F_{D_1}+\eps\hN\ilap\calL\ilap F_{\phi_1}+\eps\hN\ilap\Lambda\ilap F_{D_1}\right),
\end{eqnarray*}
we get
\begin{eqnarray*}
&& \int d^2x\int d^2x' \, \{F,\Phi_1({\bf x})\}C_{11}^{-1}({\bf x},{\bf x}')\{\Phi_1({\bf x}'),G\}=\\
&& \qquad \qquad \frac{1}{\eps}\int d^2x\,  (F_{n_1}+\ilap F_{\phi_1})(1-\hN\ilap)^{-1}\hN\ilap\calL\ilap\hN^\dagger(1-\ilap\hN^\dagger)^{-1}(G_{n_1}+\ilap G_{\phi_1}),\\
	&& \int d^2x\int d^2x' \, \{F,\Phi_1({\bf x})\}C_{12}^{-1}({\bf x},{\bf x}')\{\Phi_2({\bf x}'),G\}=\frac{1}{\eps^2} \int d^2x \, (F_{n_1}+\ilap F_{\phi_1})G_{D_1} \\
	&& \qquad \qquad -\frac{1}{\eps}\int d^2x \left((\ilap \Lambda\ilap F_{\phi_1}-\ilap\calL\ilap F_{D_1})G_{D_1}-(F_{n_1}+\ilap F_{\phi_1})(1-\hN\ilap)^{-1}\hN\ilap \calL\ilap G_{\phi_1} \right), \\
	&& \int d^2x\int d^2x' \, \{F,\Phi_2({\bf x})\}C_{21}^{-1}({\bf x},{\bf x}')\{\Phi_1({\bf x}'),G\}=-\frac{1}{\eps^2} \int d^2x \, F_{D_1}(G_{n_1}+\ilap G_{\phi_1})\\
	&& \qquad \qquad +\frac{1}{\eps}\int d^2x F_{D_1}(\ilap\Lambda\ilap G_{\phi_1}-\ilap\calL\ilap G_{D_1})- \hN\ilap\calL\ilap F_{\phi_1}(1-\ilap\hN^\dagger)^{-1}(G_{n_1}+\ilap G_{\phi_1}),\\
	&& \int d^2x\int d^2x'\, \{F,\Phi_2({\bf x})\}C_{22}^{-1}({\bf x},{\bf x}')\{\Phi_2({\bf x}'),G\}=\frac{1}{\eps}\int d^2x \, F_{D_1}\ilap\calL\ilap G_{D_1}.
\end{eqnarray*}

By summing these contributions and subtracting the result from the  Poisson bracket~(\ref{eq:PB1}),  as per (\ref{eqn:dirac}), we obtain the following Dirac bracket:
$$
\{F,G\}_*=-\frac{1}{\eps}\int d^2x \, (\Delta-\hN^\dagger)^{-1} (F_{\phi_1}+\hN^\dagger F_{n_1})\calL (\Delta -\hN^\dagger)^{-1} (G_{\phi_1}+\hN^\dagger G_{n_1}).
$$
If we define $\bar{F}$ for any functional $F[\phi_1,n_1,D_1]$ as
$$
\bar{F}[\phi_1]=F[\phi_1,N(\phi_1),0],
$$
the functional derivative of $\bar{F}$ with respect to $\phi_1$ is given by
$$
\bar{F}_{\phi_1}=F_{\phi_1}+\hN^\dagger F_{n_1}.
$$
Therefore, the Dirac bracket becomes
\begin{equation}
\label{eq:HMrpb}
\{F,G\}_*=\frac{1}{\eps}\int d^2x \, (\lap\phi_1-N(\phi_1))\left[(\Delta-\hN^\dagger)^{-1}\bar{F}_{\phi_1},(\Delta-\hN^\dagger)^{-1}\bar{G}_{\phi_1}\right].
\end{equation}
Finally, if  we perform the following change of variables 
$$
q_1=\Delta\phi_1-N(\phi_1),
$$
the Dirac bracket achieves the compact form 
\begin{equation}
\{F,G\}_*=\frac{1}{\eps}\int d^2x \, q_1[F_{q_1},G_{q_1}],
\label{qbkt}
\end{equation}
as obtained in Ref.~\cite{tass09}. In particular, we notice that the Poisson bracket~(\ref{qbkt}) does not depend explicitly on the constraint function $N$. Therefore,  all the models with constraints of the form $n_1=N(\phi_1)$ share the same Poisson bracket when expressed in terms of the generalized vorticity variable $q_1$. 



\section{Hamiltonians}  \label{sec:hamilt}

Having obtained a common Poisson bracket for our class of systems, we now obtain  Hamiltonians for various cases. 


\subsection{2D Euler equation}

When the  electrostatic potential is set to zero, the expansion of Hamiltonian~(\ref{eq:hamnv}) is given by
$$
H_1[n_1,\vv_1]=\frac{\eps}{2} \int d^2 x \, {v_1^2}\,, 
$$
which in terms of the variables $\phi_1$ and $D_1$ is 
$$
H_1=\frac{\eps}{2} \int d^2x  \, \left(|\nabla \phi_1|^2+|\nabla D_1|^2\right)\,.
$$
As constraints we choose a constant density, i.e.,  $N=0$, and   incompressibility, i.e.,   $D_1=0$.  The resulting Dirac bracket is given by
$$
\{F,G\}_*=\frac{1}{\eps}\int d^2x\,  \lap \phi_1 [\ilap \bar{F}_{\phi_1},\ilap \bar{G}_{\phi_1}].
$$
This corresponds to the Hamiltonian and the Poisson bracket of the two-dimensional Euler equation, which was given  in \cite{morr82,olver82}  in terms of the vorticity variable $q_1=\lap \phi_1$:
$$
H_1=-\frac{1}{2}\int d^2 x \, q_1 \ilap q_1,
$$
and 
$$
\{F,G\}_*=\int d^2x \, q_1 [F_{q_1},G_{q_1}]\,.
$$
 Since $\eps$ does no longer play a role in the discussion, we drop it in what follows. 


\subsection{Charney-Hasegawa-Mima equation}

Now we assume a linear adiabatic response of the electrons. 
Upon expanding $\varphi(n)=\log(n/(1-\eps \lambda))$,   this gives 
$N(\phi_1)=\phi_1-\lambda$ and $ \hN=1$.
The Dirac bracket becomes
$$
\{F,G\}_*=  \int d^2x \, (\lap\phi_1-\phi_1+\lambda)\left[(\Delta-1)^{-1}\bar{F}_{\phi_1},(\Delta-1)^{-1}\bar{G}_{\phi_1}\right],
$$
which is exactly the Poisson bracket found in Ref.~\cite{wein83} and derived in Ref.~\cite{tass09}.
The expansion of Hamiltonian~(\ref{eq:hamnv}) is given by~\cite{tass09} 
$$
H_1=\frac{1}{2} \int d^2x \, \left(\vert\nabla\phi_1\vert^2+\vert\nabla D_1\vert^2+n_1^2+2\lambda n_1 \right),
$$
which becomes
$$
H_1=\frac1{2}\int d^2x \, \left(\vert \nabla \phi_1\vert^2+\phi_1^2\right)\,, 
$$
when introducing the constraints $n_1=N(\phi_1)$ and $D_1=0$.

We notice that for $N(\phi_1)= \phi_1/\eta$ where $\eta$ is small, we recover the Hamiltonian structure of the asymptotic model given in Ref.~\cite{lari91}.  


\subsection{General setting}

We consider a Hamiltonian which is given by the sum of a kinetic energy plus a potential part
\begin{equation}
\label{eq:hamG}
H=\int d^2 x \, \frac{\vert \nabla \phi_1\vert^2}{2} + V[N(\phi_1)],
\end{equation}
where 
\begin{equation}
V[n]=\int d^2x\,  \psi(n)\,,
\label{eq:V}
\end{equation}
 is a quite general functional of the density $n$ and $\psi$ is arbitrary.  With this Hamiltonian and the bracket of (\ref{eq:HMrpb}) or (\ref{qbkt}), the  equation of motion for $\phi_1$ is given by
$$
(\Delta-\hN)\dot{\phi}_1=-\left[\Delta\phi_1-N(\phi_1),(\Delta -\hN^\dagger)^{-1}\bar{H}_{\phi_1}\right],
$$
where 
$$
\bar{H}_{\phi_1}=-\lap \phi_1 +\hN^\dagger V_n.
$$
This equation which originates from a Hamiltonian system is valid for any constraint $N$ and any potential $V$.  As stated above, all these models share the same Poisson bracket (in the generalized vorticity variable). They differ in their Hamiltonian.  For $V$ given by Eq.~(\ref{eq:V}) and for $\psi$ such that
$$
\psi'(N(\phi_1))=\phi_1,
$$
this equation simplifies,  
$$
(\Delta-\hN)\dot{\phi}_1=\left[\Delta\phi_1-N(\phi_1),\phi_1\right],
$$
which is verified for the two cases described above, the 2D Euler and the CHM equations. As a consequence of our ordering, this choice of $\psi$ is equivalent to the condition mentioned in Sec.~\ref{sec:ion}, namely that $\psi'(n)=\varphi(n)$.

The question now becomes whether or not it is possible to construct a  potential $V$ for a dynamical equation or an  electron response of interest.  A rigorous answer is determined for the dynamical equation for the electric potential $\varphi$ generated by the electrons.  For example, in the  nonlinear drift wave  model derived by Terry and Horton  \cite{terr82} (see also \cite{hort86,hort99}), the electric potential is not only a function of $n$ but of all its derivatives, namely, 
$$
\varphi(n)=N^{-1}(n), 
$$
where $N^{-1}$ is the inverse of the relation $n=N(\phi)$. In particular,  $N(\varphi)=(1+{\mathcal O}^{\rm a})\varphi-\lambda$,  where ${\mathcal O}^{\rm a}=\delta_0 (c_1+\lap)\partial_y$ is an anti-self-adjoint operator.  For such cases, the potential function $\psi$ with $\psi'=\varphi$ cannot be constructed.  General potential functions of the form of that of (\ref{eq:hamG}), such that
$$
V_n=\varphi(n) 
$$
can {\it only} be solved  if $\hN$ is a self-adjoint operator.  Thus the Terry-Horton model is not a Hamiltonian model in our class.



\section{Conclusions} \label{sec:concl}

We have shown how the  theory of constrained Hamiltonian systems, developed by Dirac, provides an effective way to include constraints into a Hamiltonian system while preserving the  Hamiltonian character. In particular,  we have applied this theory to a case relevant for plasma physics, by adopting as the parent model a fluid system consisting of the ion momentum equation and   the continuity equation, where the  electron dynamics determines the electrostatic potential as a function of the density. We derived the Poisson bracket for the Euler and the Charney-Hasegawa-Mima equations as the Dirac bracket of this dynamics,  obtained by considering an incompressibility condition and a general relation between the density and the velocity field as constraints.  Given  the bracket thus obtained, different choices for the Hamiltonian functional lead to different models, and in this way a general class of systems was derived, all of which preserve the constraints and are Hamiltonian by construction. 

We  emphasize that the technique presented here is general, and can be applied   to derive new reduced Hamiltonian models  from more general parent models, models for which the  Hamiltonian structure becomes directly available.

\acknowledgments
This work was supported by the European Community under the contract of Association
between EURATOM, CEA, and the French Research Federation for fusion studies. The views and opinions expressed herein do not necessarily reflect those of the European Commission. Financial support was also received from the Agence Nationale de la Recherche (ANR EGYPT) and PJM was supported by the US Dept.\  of Energy Contract  DE-FG03-96ER-54346.

\end{document}